\documentclass{jetpl}
\usepackage{color,amsmath,amsfonts,graphics,epsfig,amssymb}
\twocolumn

\lat

\title{
Search for Galactic disk and halo components in the arrival
directions of high-energy astrophysical neutrinos. }

\rtitle{Arrival directions of high-energy neutrinos}

\sodtitle{
Search for Galactic disk and halo components in the arrival
directions of high-energy astrophysical neutrinos. }

\author{S.\,V.\,Troitsky\thanks{e-mail: st@ms2.inr.ac.ru}}

\rauthor{S.\,V.\,Troitsky}

\sodauthor{S.\,V.\,Troitsky}

\address{Institute for Nuclear Research of the Russian Academy of
Sciences,\\
60th October Anniversary prospect 7A, 117312 Moscow, Russia}

\dates{November 4, 2015}{*}

\abstract{
Arrival directions of 40 neutrino events with energies $\gtrsim 100$~TeV,
observed by the IceCube experiment, are studied. Their distribution in the
Galactic latitude and in the angular distance to the Galactic Center
allow to search for the Milky-Way disk and halo-related
components, respectively. No statistically significant evidence for the
disk component is found, though even
100\% disk origin of the flux is allowed at the 90\% confidence level.
Contrary, the Galactic Center--Anticenter dipole anisotropy, specific for
dark-matter decays (annihilation) or for interactions of cosmic rays with
the extended halo of circumgalactic gas, is clearly favoured over the
isotropic distribution (the probability of a fluctuation of the isotropic
signal is $\sim 2\%$). }


\begin{document}

\maketitle

The origin of high-energy astrophysical neutrinos, discovered recently by
the IceCube experiment \cite{IceCube1, IceCube2, HESE-3yr, muon-events},
is
unknown. Large flux, relatively soft spectrum and lack of anisotropy in
arrival directions make it difficult to explain the observed flux in terms
of a single population of well-understood sources. It is even unknown,
which fraction of the flux may come from Galactic sources. Some
concentration of the high-energy starting events (HESE) with energies $E
\gtrsim 60$~TeV towards the Galactic Center, visible in early skymaps,
suggested a possible significant Galactic fraction. This may be related to
a particular population of sources in the Milky Way, to interactions of
cosmic rays with the interstellar matter, to the same process in the
circumgalactic gas halo, or to decay or annihilation of dark-matter
particles in the Galactic dark halo. The former two scenarios should
reveal themselves via excess of events coming from the Galactic plane,
while the latter two explanations imply the Galactic center-anticenter
asymmetry due to the non-central position of the Solar System in
the Galaxy. A recent study \cite{NerSemGal} claimed an evidence for the
disk component in the distribution of arrival directions of 19 HESE
neutrinos, observed by IceCube in four years, with estimated energies
above 100~TeV, where the contribution of the atmospheric background events
is minimal. The aim of the present study is to search for the Galactic
component in the combined sample of these 19 events and of 21 high-energy
muon-neutrino events (HEmu) in the similar energy range from the analysis
of Ref.~\cite{muon-events}. In what follows, I compare the observed
distributions of arrival directions in the Galactic latitude $b$ (test of
the disk component) and in the angular distance to the Galactic Center
$\Theta$ (test of the halo scenarios) with similar Monte-Carlo
distributions combining the (known) background contribution with the
assumed mixture of the Galactic and isotropic components in the signal.

\textit{\textbf{Data. ---}}
The data set studied here consists of two parts.
The first is the 4-year HESE sample whose arrival directions are published
in Refs.~\cite{HESE-3yr, HESE-4yr}. From the list of 54 events, 19
neutrinos with estimated energies above 100~TeV were selected. The
choice of the energy cut is rather arbitrary but is motivated by two
facts. First, above this or similar energy, the contribution of background
atmospheric neutrinos becomes small compared to the presumably
astrophysical flux. Indeed, the mean expected number of background events
is 2.26, as one may conclude from Fig.~3 of Ref.~\cite{HESE-4yr}. Second,
arrival directions of events of these energies are published for the
second sample, so the use of the energy threshold makes it possible to
consider the largest sample of high-energy events jointly.

The second part is the sample of muon neutrino tracks from the Northern
sky \cite{muon-events}, whose arrival directions are published in a public
data release of the IceCube collaboration \cite{muon-events-list}. There
are 21 such events; their estimated energies, quoted in the
catalog~\cite{muon-events-list}, are above 100~TeV. One of the events in
the sample is present in the HESE sample as well; however, its
reconstructed energy in the HESE analysis is below 100~TeV, so it is not
counted twice. From Fig.~2 of Ref.~\cite{muon-events}, one concludes that
the mean expected number of background events in the sample is a sum of
6.15 conventional atmospheric neutrinos and 0.86 prompt atmospheric
neutrinos. Therefore, the data set used here contains 40 events with the
mean expected background of 9.27 events.

\textit{\textbf{Monte-Carlo set: background. ---}}
For the HESE sample, the expected distribution of background events in the
zenith angle is presented in Supplementary Figure~5 of
Ref.~\cite{HESE-3yr}. Note that for the South-Pole location of IceCube,
the zenith angle is uniquely translated to declination, which allows to
generate Monte-Carlo distribution of background events easily.

For the HEmu sample, one considers two components of the background
separately. The contribution of prompt atmospheric neutrinos, not
negligible at relevant high energies, is isotropic. To obtain the
distribution of conventional atmospheric neutrinos in the zenith angle,
I follow Refs.~\cite{Volkova, Chirkin}. Both distributions are convolved
with the zenith-angle-dependent acceptance of the experiment which may be
read from Fig.~3b of Ref.~\cite{1408.0634} (see also Fig.~6.2b of
Ref.~\cite{thesis}). The acceptance for a sample of 0.1\% most energetic
events is used; according to Table~6.1 of Ref.~\cite{thesis}, this roughly
corresponds to energies above $\sim 120$~TeV.

\textit{\textbf{Monte-Carlo set: isotropic signal. ---}}
The assumed isotropic signal is modified by the detector acceptance, which
is described in the previous paragraph for the HEmu sample. For HESE, the
simulated distribution of isotropic events in zenith angle at $E>100$~TeV
is also given in Supplementary Figure~5 of Ref.~\cite{HESE-3yr}.

\textit{\textbf{Monte-Carlo set: Galactic-disk signal. ---}}
The notion of the Galactic-disk signal is ambiguous, since various
populations of potential sources follow different distributions. Since the
neutrino emission is accompanied by gamma rays in most scenarios, the
distribution of the Galactic diffuse gamma-ray flux is often used as a
template for the disk neutrino flux, see
e.g.\ Refs.~\cite{1412.1690, Ahlers}. Here, we use the FERMI--LAT template
of Ref.~\cite{1202.4039}, for the HI contribution at gamma-ray energies
(0.2--1.6)~GeV (the dependence on details of the template chosen is beyond
the overall precision of the analysis). The distribution of arrival
directions in this and further cases is corrected for the detector
acceptance, as described above, for the HESE and HEmu samples separately.

\textit{\textbf{Monte-Carlo set: dark-matter signal. ---}}
To emulate the anisotropy pattern of the neutrino signal from
dark-matter decays, we use explicit expressions of Ref.~\cite{Khrenov} and
assume the Burkert \cite{Burkert} distribution of dark matter.
This allows one to obtain the distribution of arrival directions in the
angular distance to the Galactic Center (see Ref.~\cite{esmaili} for a
study of early $E>60$~TeV IceCube data). For the case of dark-matter
annihilation, the density of dark matter $n$ is replaced by $n^{2}$ in the
same expressions.

\textit{\textbf{Monte-Carlo set: outer-halo signal. ---}}
Following Refs.~\cite{Ahar-gas, OK-ST-gamma, OK-ST-progress}, we consider
the possibility that neutrinos are produced by cosmic-ray interactions
with gas in the extended (up to $\sim 200$~kpc) outer halo of the Galaxy.
For the target-gas distribution, we use that of Ref.~\cite{1412.3116},
$$
n_{\rm gas} \propto \left(1+\left(r/r_{\rm c} \right)^2   \right)^{-3
\beta/2},
$$
with $\beta=0.5$ and $r_{\rm c}
=5$~kpc~\cite{1412.3116}. For the
cosmic-ray distribution, we use $n_{\rm CR}\propto 1/(1+r/r_{1})$ with
$r_{1}=20$~kpc to reproduce the asymptotic used in Ref.~\cite{Ahar-gas}.
The product $n_{\rm gas}n_{\rm CR}$ replaces $n$ in expressions of
Ref.~\cite{Khrenov}.

\begin{figure}
\includegraphics[width=0.9\columnwidth]{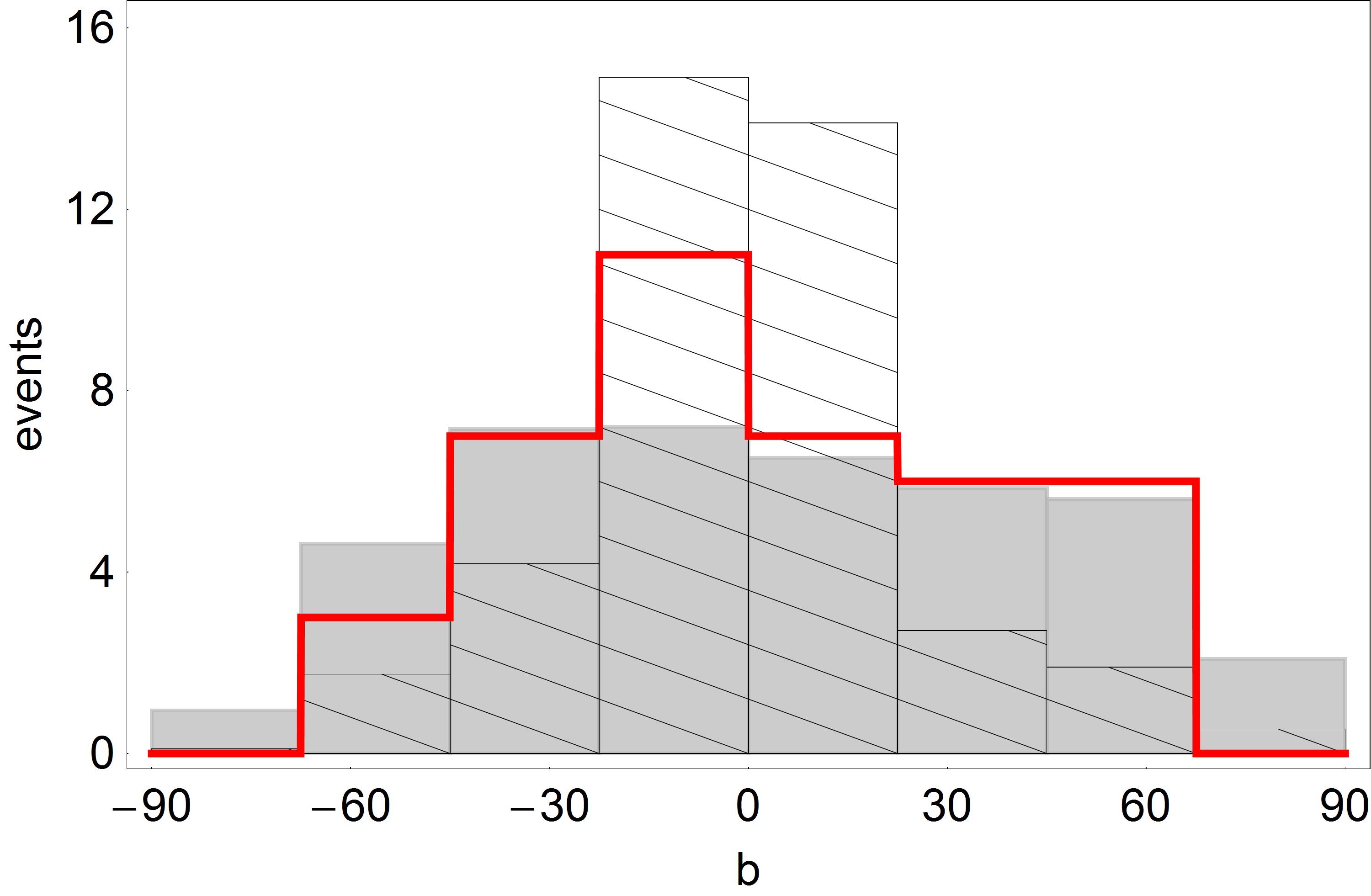}
\caption{\label{fig:histogram-disk}
Figure~\ref{fig:histogram-disk}.
Distribution of arrival directions in galactic latitude $b$. Full line
(red online) -- data; shaded histogram -- background plus isotropic signal;
hatched histogram -- background plus Galactic disk signal. }
\end{figure}

\begin{figure}
\includegraphics[width=0.9\columnwidth]{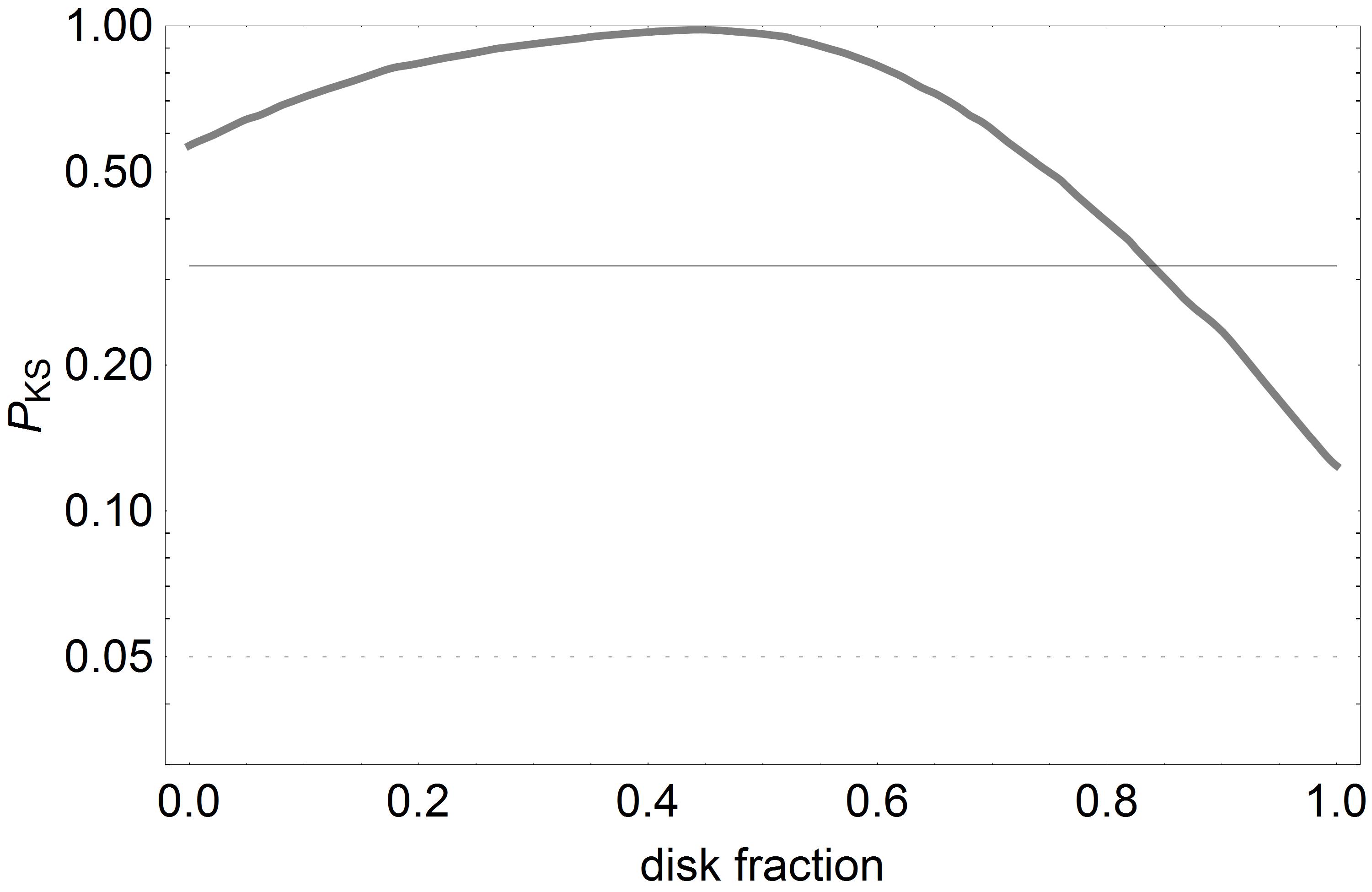}
\caption{\label{fig:P-disk}
Figure~\ref{fig:P-disk}.
The Kolmogorov-Smirnov probability $P_{\rm KS}$ that the observed
distribution of events in the Galactic latitude $b$ is a fluctuation of
a model distribution in which the signal is a mixture of the disk
fraction $\xi_{\rm d}$ and the isotropic fraction $(1-\xi_{\rm d})$, as a
function of $\xi_{\rm d}$ (the full grey curve). Horizontal lines
indicate $1-P_{\rm KS}=0.68$ (full) and 0.95 (dotted): the values of
$\xi_{\rm d}$ for which the curve is below the lines are excluded at the
68\% and 95\% CL, respectively. }
\end{figure}

\textit{\textbf{Results. ---}}
Figure~\ref{fig:histogram-disk}
presents the distribution of simulated and
observed events in the Galactic latitude. Composing the simulated data
set of a known number of atmospheric background events, a
fraction of $\xi_{\rm d}$ from the disk component and $(1-\xi_{\rm d})$
from the isotropic distribution, it is easy to compare the distribution of
observed and simulated events by means of the Kolmogorov-Smirnov test,
which gives the probability $P_{\rm KS}$ that the observed distribution is a
statistical fluctuation of the simulated one, see Fig.~\ref{fig:P-disk}.
The isotropic distribution is perfectly consistent with data ($P_{\rm
KS} \approx 0.57$). However, as one can see from Fig.~\ref{fig:P-disk}, all
values, $0 \le \xi_{\rm d} \le 1$, are allowed with $P_{\rm KS}>0.1$, that
is at least at the 90\% confidence level (CL).

\begin{figure}
\includegraphics[width=0.9\columnwidth]{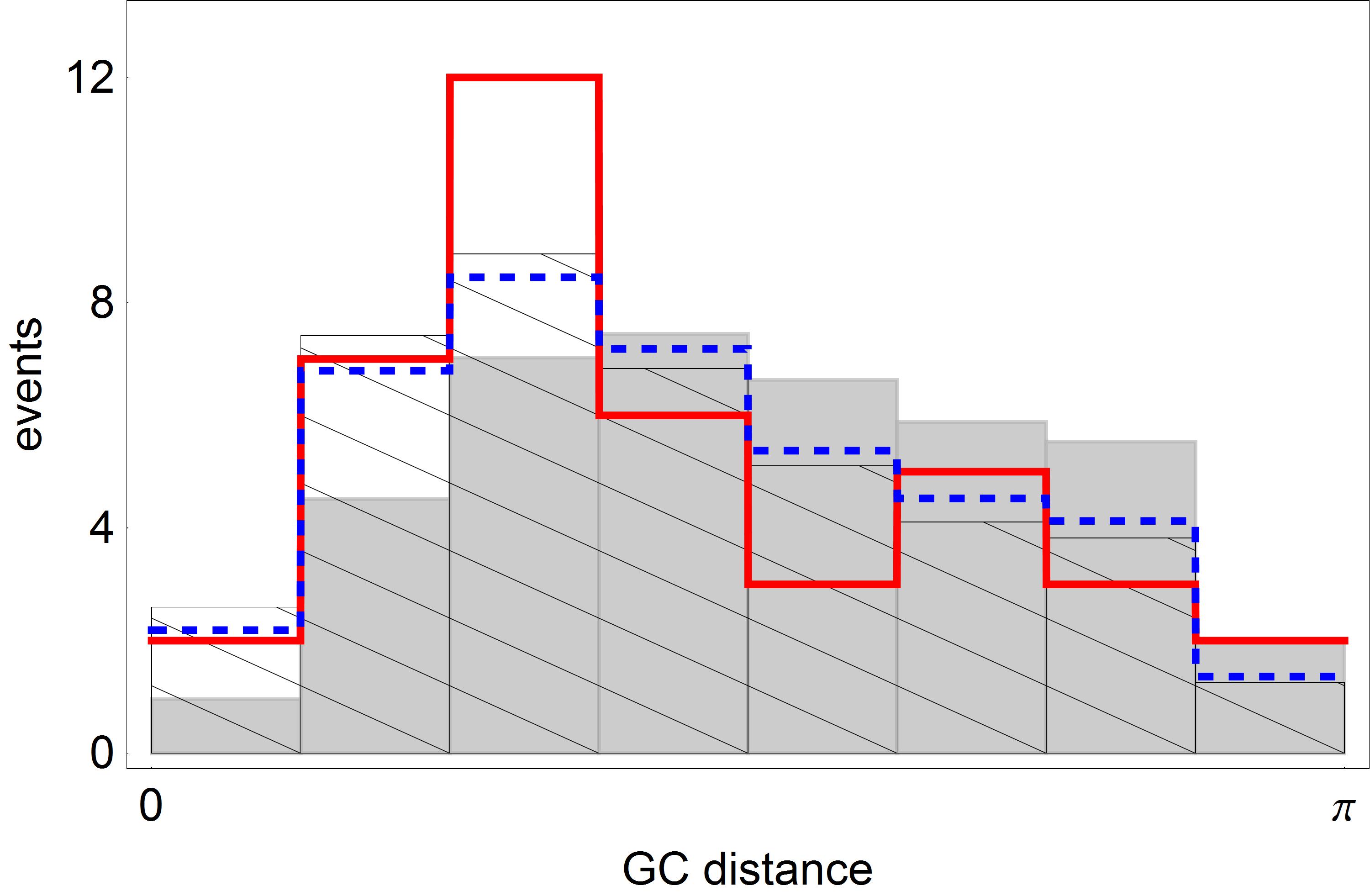}
\caption{\label{fig:histogram-annih}
Figure~\ref{fig:histogram-annih}.
Distribution of arrival directions in the angular distance $\Theta$ to the
Galactic Center. Full line (red online) -- data; shaded histogram --
background plus isotropic signal; hatched histogram -- background
plus signal from dark-matter annihilation in the Milky Way; dashed line
(blue online) -- background
plus signal from dark-matter decays in the Milky Way. }
\end{figure}

\begin{figure}
\includegraphics[width=0.9\columnwidth]{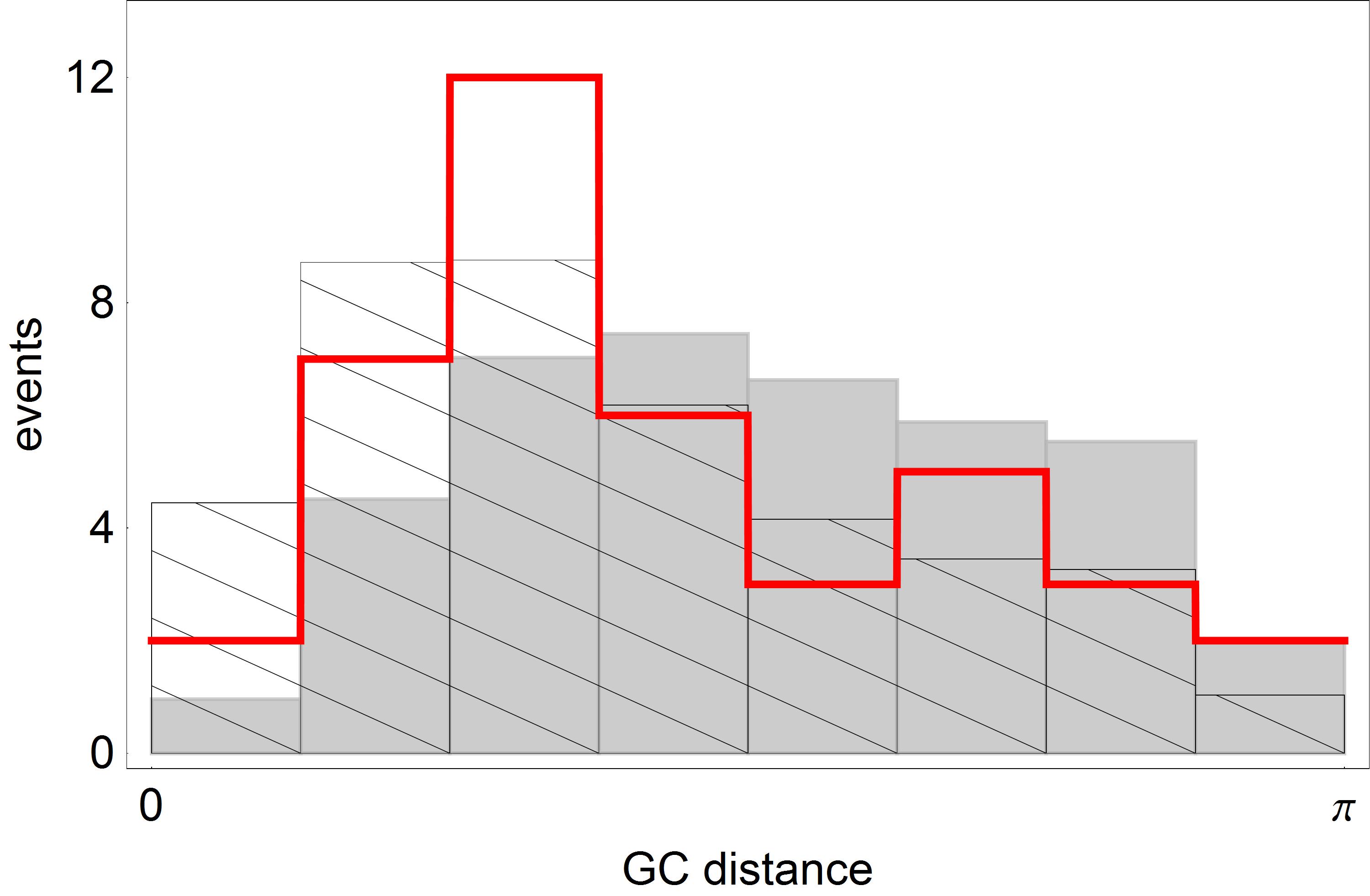}
\caption{\label{fig:histogram-gas}
Figure~\ref{fig:histogram-gas}.
Distribution of arrival directions in the angular distance to the
Galactic Center. Full line (red online) -- data; shaded histogram --
background plus isotropic signal; hatched histogram -- background
plus signal from cosmic-ray interactions with the halo of circumgalactic
gas. }
\end{figure}

\begin{figure}
\includegraphics[width=0.9\columnwidth]{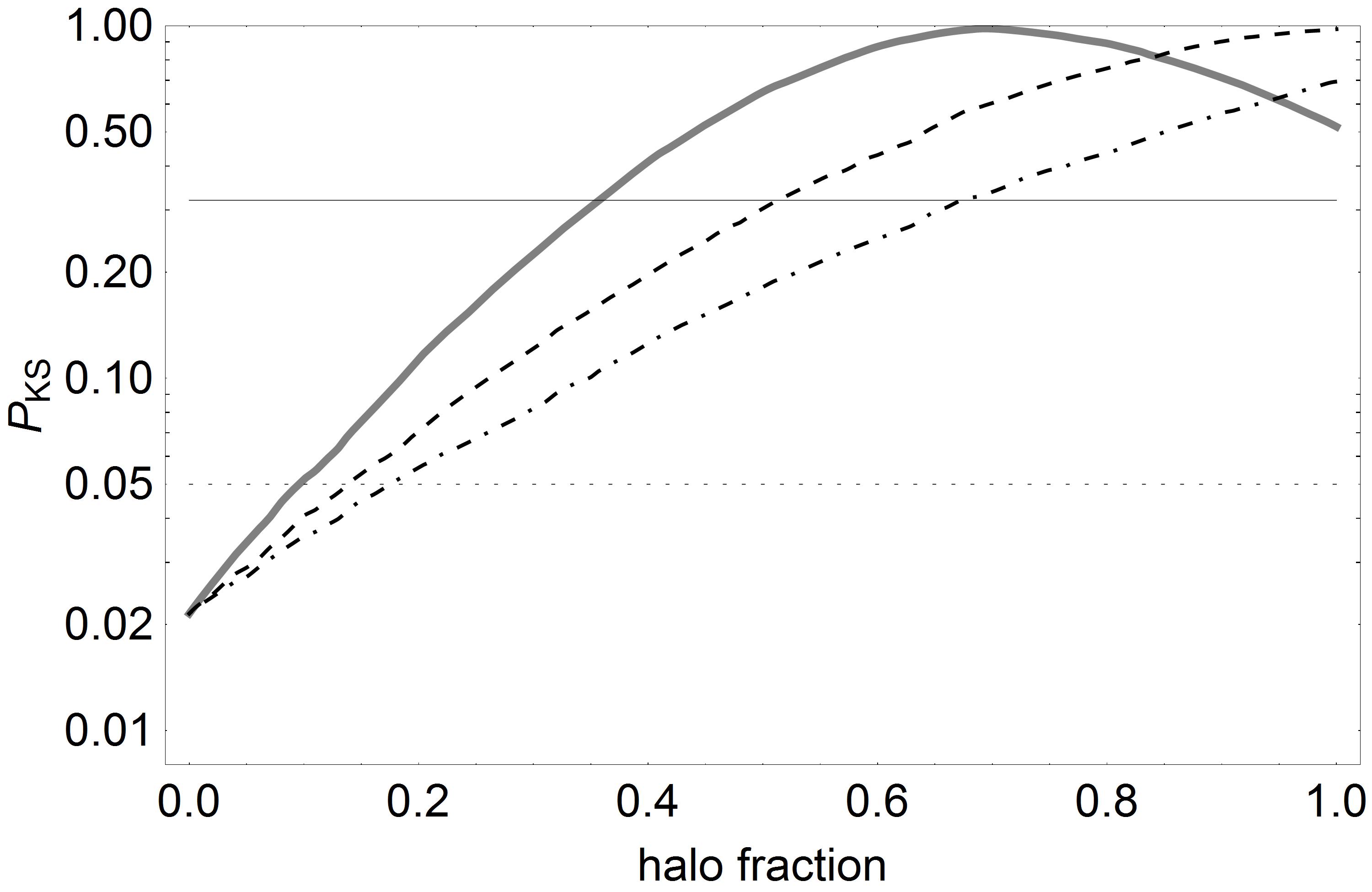}
\caption{\label{fig:P-decay}
Figure~\ref{fig:P-decay}.
The Kolmogorov-Smirnov probability $P_{\rm KS}$ that the observed
distribution of events in the angular distance to the Galactic Center is a
fluctuation of a model distribution in which the signal is a mixture of
the fraction $\xi_{\rm h}$ coming from halo and the
remaining fraction $(1-\xi_{\rm h})$ isotropic, as a function of $\xi_{\rm
h}$ (full grey curve: cosmic-ray interactions with circumgalactic gas;
dashed curve: dark-matter annihilation; dash-dotted curve: dark matter
decays). Horizontal lines indicate $1-P_{\rm KS}=0.68$ (full) and
0.95 (dotted): the values of $\xi_{\rm h}$ for which a curve is below the
lines are excluded at the 68\% and 95\% confidence level, respectively. }
\end{figure}

For the halo scenarios, a similar analysis was performed in terms of the
angular distance $\Theta$ between the arrival direction and the Galactic
Center. The distributions of data and simulated event sets in $\Theta$ are
shown in Figs.~\ref{fig:histogram-annih}, \ref{fig:histogram-gas}.
The data favours the dipole anisotropy, either in the dark-matter decay or
in the circumgalactic gas halo scenario, over isotropy (see
Fig.~\ref{fig:P-decay}).
For the isotropic distribution, $P_{\rm KS}\approx
0.02$, while $P_{\rm KS} >0.5$ for all three pure
halo scenarios.

To summarize, the sample of 40 IceCube events with $E \gtrsim 100$~TeV, of
which $\sim 9$ are background, neither shows a statistically
significant evidence for, nor exclude, the Galactic disk component. The
Galactic Center-Anticenter dipole, contrary, is favoured over isotropy at
98\% CL, which may be a signal of the Galactic halo component related
either to dark-matter decays (annihilation) or to cosmic-ray interactions
with circumgalactic gas.
Further studies of high-energy neutrinos are mandatory to
make stronger conclusions. In particular, more uniform full-sky statistics
is important for global anisotropy studies, and will be provided in coming
years with joint efforts of the South-Pole IceCube and Northern-hemisphere
experiments: Baikal-GVD \cite{Baikal1, Baikal2, Baikal3}, whose first
cluster is taking data since April 2015 and further ones are to be
deployed next winter, and KM3NET \cite{KM3NET} whose construction is
expected to start.

\textit{\textbf{Acknowledgements. ---}}
I thank O.~Kalashev, M.~Pshirkov and G.~Rubtsov for
discussions, D.~Semikoz for useful critical comments on the manuscript and
CERN (PH-TH unit) for hospitality at the initial stages of this work. This
work was supported by the Russian Science Foundation, grant 14-12-01340.

\end{document}